\documentclass[%
 reprint,
superscriptaddress,
 amsmath,amssymb,
 aps,
prl,
]{revtex4-2}

\usepackage{graphicx}
\usepackage{dcolumn}
\usepackage{bm}
\usepackage{xcolor}
\usepackage{float}	
\usepackage{soul}
\usepackage{url}
\begin{document}

\newcommand{\bra}[1]{\left\langle\,#1\,\right|}
\newcommand{\ket}[1]{\left|\,#1\,\right\rangle}

\preprint{APS/123-QED}

\title{Chirality-Driven Orbital Angular Momentum and Circular Dichroism in CoSi}

\author {Stefanie~Suzanne~Brinkman}
\email{stefanie.brinkman@ntnu.no}
\affiliation {Center for Quantum Spintronics, Department of Physics, Norwegian University of Science and Technology, 7491 Trondheim, Norway}

\author {Xin~Liang~Tan}
\affiliation {Center for Quantum Spintronics, Department of Physics, Norwegian University of Science and Technology, 7491 Trondheim, Norway}
\affiliation {Peter Gr\"{u}nberg Institut (PGI-6), Forschungszentrum J\"{u}lich, J\"{u}lich 52425, Germany}
\affiliation {Faculty of Physics, University of Duisburg-Essen, Duisburg 47057, Germany}

\author {Bj\o rnulf~Brekke}
\affiliation {Center for Quantum Spintronics, Department of Physics, Norwegian University of Science and Technology, 7491 Trondheim, Norway}

\author {Anders~Christian~Mathisen}
\affiliation {Center for Quantum Spintronics, Department of Physics, Norwegian University of Science and Technology, 7491 Trondheim, Norway}

\author {\O yvind~Finnseth}
\affiliation {Department of Materials Science and Engineering, Norwegian University of Science and Technology, 7491 Trondheim, Norway}

\author {Richard Justin Schenk}
\affiliation {Center for Quantum Spintronics, Department of Physics, Norwegian University of Science and Technology, 7491 Trondheim, Norway}

\author {Kenta~Hagiwara}
\affiliation {Peter Gr\"{u}nberg Institut (PGI-6), Forschungszentrum J\"{u}lich, J\"{u}lich 52425, Germany}
\affiliation {Faculty of Physics, University of Duisburg-Essen, Duisburg 47057, Germany}

\author{Meng-Jie Huang}
\affiliation{Ruprecht Haensel Laboratory, Kiel University and DESY, Germany}
\affiliation{Deutsches Elektronen-Synchrotron DESY, Notkestra{\ss}e 85, 22607 Hamburg, Germany}

\author{Jens Buck}
\affiliation{Institut f\"{u}r Experimentelle und Angewandte Physik, Christian-Albrechts-Universit\"{a}t zu Kiel, 24098 Kiel, Germany}
\affiliation{Ruprecht Haensel Laboratory, Kiel University and DESY, Germany}

\author{Matthias Kall\"{a}ne}
\affiliation{Institut f\"{u}r Experimentelle und Angewandte Physik, Christian-Albrechts-Universit\"{a}t zu Kiel, 24098 Kiel, Germany}
\affiliation{Ruprecht Haensel Laboratory, Kiel University and DESY, Germany}

\author{Moritz Hoesch}
\affiliation{Deutsches Elektronen-Synchrotron DESY, Notkestra{\ss}e 85, 22607 Hamburg, Germany}

\author{Kai Rossnagel}
\affiliation{Institut f\"{u}r Experimentelle und Angewandte Physik, Christian-Albrechts-Universit\"{a}t zu Kiel, 24098 Kiel, Germany}
\affiliation{Ruprecht Haensel Laboratory, Kiel University and DESY, Germany}
\affiliation{Deutsches Elektronen-Synchrotron DESY, Notkestra{\ss}e 85, 22607 Hamburg, Germany}

\author{Kui-Hon Ou Yang}
\affiliation {Department of Physics, National Taiwan University, Taipei 10617, Taiwan}

\author{Minn-Tsong Lin}
\affiliation {Department of Physics, National Taiwan University, Taipei 10617, Taiwan}
\affiliation{Institute of Atomic and Molecular Sciences, Academia Sinica, Taipei
10617, Taiwan}
\affiliation{Research Center for Applied Sciences, Academia Sinica, Taipei 11529,
Taiwan}

\author{Guo-Jiun Shu} 
\affiliation {Department of Materials and Mineral Resources Engineering, National
Taipei University of Technology, Taipei, Taiwan}

\author {Ying-Jiun~Chen}
\affiliation {Peter Gr\"{u}nberg Institut (PGI-6), Forschungszentrum J\"{u}lich, J\"{u}lich 52425, Germany}
\affiliation {Faculty of Physics, University of Duisburg-Essen, Duisburg 47057, Germany}
\affiliation {Ernst Ruska-Centre for Microscopy and Spectroscopy with Electrons and
Peter Gr\"{u}nberg Institute, Forschungszentrum J\"{u}lich, 52425 J\"{u}lich, Germany}

\author {Christian~Tusche}
\affiliation {Peter Gr\"{u}nberg Institut (PGI-6), Forschungszentrum J\"{u}lich, J\"{u}lich 52425, Germany}
\affiliation {Faculty of Physics, University of Duisburg-Essen, Duisburg 47057, Germany}

\author {Hendrik~Bentmann}
\affiliation {Center for Quantum Spintronics, Department of Physics, Norwegian University of Science and Technology, 7491 Trondheim, Norway}

\date{\today}

\begin{abstract} 
Chiral crystals and molecules were recently predicted to form an intriguing platform for unconventional orbital physics. Here, we report the observation of chirality-driven orbital textures in the bulk electronic structure of CoSi, a prototype member of the cubic B20 family of chiral crystals. Using circular dichroism in soft X-ray angle-resolved photoemission, we demonstrate the formation of a bulk orbital-angular-momentum texture and monopole-like orbital-momentum locking that depends on crystal handedness. We introduce the intrinsic chiral circular dichroism, \textit{ic}CD, as a differential photoemission observable and a natural probe of chiral electron states. Our findings render chiral crystals promising for spin-orbitronics applications.
\end{abstract}

\maketitle 
Quantum phenomena resulting from structural chirality are attracting growing interest in condensed matter physics and related disciplines. Structural chirality refers to an inherent asymmetry in the arrangement of atoms within a crystal lattice or molecule, breaking spatial inversion symmetry and all other improper symmetry operations such as reflection \cite{fecher2022}. Chiral crystals are reported to host topological band nodes in their electronic structure, associated with long surface Fermi arcs \cite{Chang2018,Rao2019,Sanchez2019,Schroeter2019} and unconventional spin textures \cite{Sakano:20,Gatti20,krieger2022}. Moreover, chirality-induced spin-selectivity (CISS) in electron transmission and transport has been observed in a variety of molecular, crystalline and nanoscale systems \cite{naaman2020,Inui:20,calavalle2022}. Recent theoretical works propose that structural chirality induces electronic chirality via orbital degrees of freedom, manifesting in bulk orbital-angular-momentum (OAM) textures with distinctive features compared to non-centrosymmetric but achiral systems \cite{liu:2021,Kim_2023}, including monopole-like orbital-momentum locking \cite{Yang2023}. Such chirality-driven orbital textures are expected to have the potential for generating large orbital Hall effects \cite{Yang2023} and to underlie CISS phenomena \cite{liu:2021,Kim_2023}, making them relevant to applications in spin-orbitronics \cite{choi2023}. 

Orbital textures in two- and three-dimensional systems have been studied successfully by linear and circular dichroism (CD) in angle-resolved photoelectron spectroscopy (ARPES) \cite{Park:2012,Kim:12,bahramy2012,cao2013,Zhu:13,Min:19,Uenzelmann:20,Beaulieu:21,beaulieu:2021,Vidal:21,Unzelmann2021,Schuler:22}. In particular, it has been shown that the bulk OAM texture, as probed by CD-ARPES, can directly reflect the monopole topological charge at band nodes and thus carries information on topological properties \cite{Unzelmann2021}. However, for systems with bulk structural chirality, the CD-ARPES response and orbital textures remain largely unexplored in experiment. Here, we focus on CoSi as a prototype chiral crystal in the B20 compound family that provides an excellent platform for investigating chirality-induced orbital textures \cite{Yang2023}. In particular, CoSi is non-magnetic and has a weak spin-orbit coupling. Consequently, structural chirality is expected to be the driving force of a finite OAM texture.

In this work, we report the observation of chirality-driven orbital momentum-locking in the bulk electronic structure of CoSi using CD-ARPES at bulk-sensitive soft X-ray excitation energies. The measured CD textures differ profoundly from previous experiments on achiral systems, and they reflect the lack of mirror symmetry in the chiral crystal structure shaping the orbital texture. We show how this unconventional CD response provides evidence for a monopole-like OAM texture, in agreement with theoretical predictions \cite{Yang2023}. CD measurements for both enantiomers of CoSi demonstrate a reversal of the OAM texture upon changing the crystal handedness. Finally, we introduce the intrinsic chiral circular dichroism \textit{ic}CD as a differential photoemission observable, to highlight the chirality-induced OAM texture.

We performed CD-ARPES experiments at the ASPHERE III endstation of the Variable Polarization XUV beamline P04 at PETRA III, DESY, Hamburg \cite{VIEFHAUS2013151}. Clean (001) surfaces of CoSi enantiomers A and B single crystals, synthesized using the  Czochralski method \cite{Marshall1968}, were prepared in situ via cycles of $\mathrm{Ar}^+$ sputtering and annealing at $\sim$950~K. Throughout the photoemission experiments, the samples were kept at a temperature of around 80~K and in an ultrahigh vacuum below $5 \times 10^{-10}$~mbar. The energy and angular resolutions of the electron analyzer were set to $<40$~meV and $<0.1$\textdegree, respectively. We resolved the out-of-plane $k_\perp$ component by collecting ARPES data using a series of photon energies between $h\nu =$~330 - 650~eV. Moreover, by considering the low-energy electron diffraction and surface Fermi arcs in ARPES as studied previously \cite{duepublico_mods_00076864, Rao2019}, we ensured the crystallographic orientation of the two enantiomers. Further details on sample growth and characterization, and out-of-plane momentum determination are described in the Supplemental Material \cite{suppl}.

\begin{figure}[t!]
    \centering
    \includegraphics[width=\linewidth]{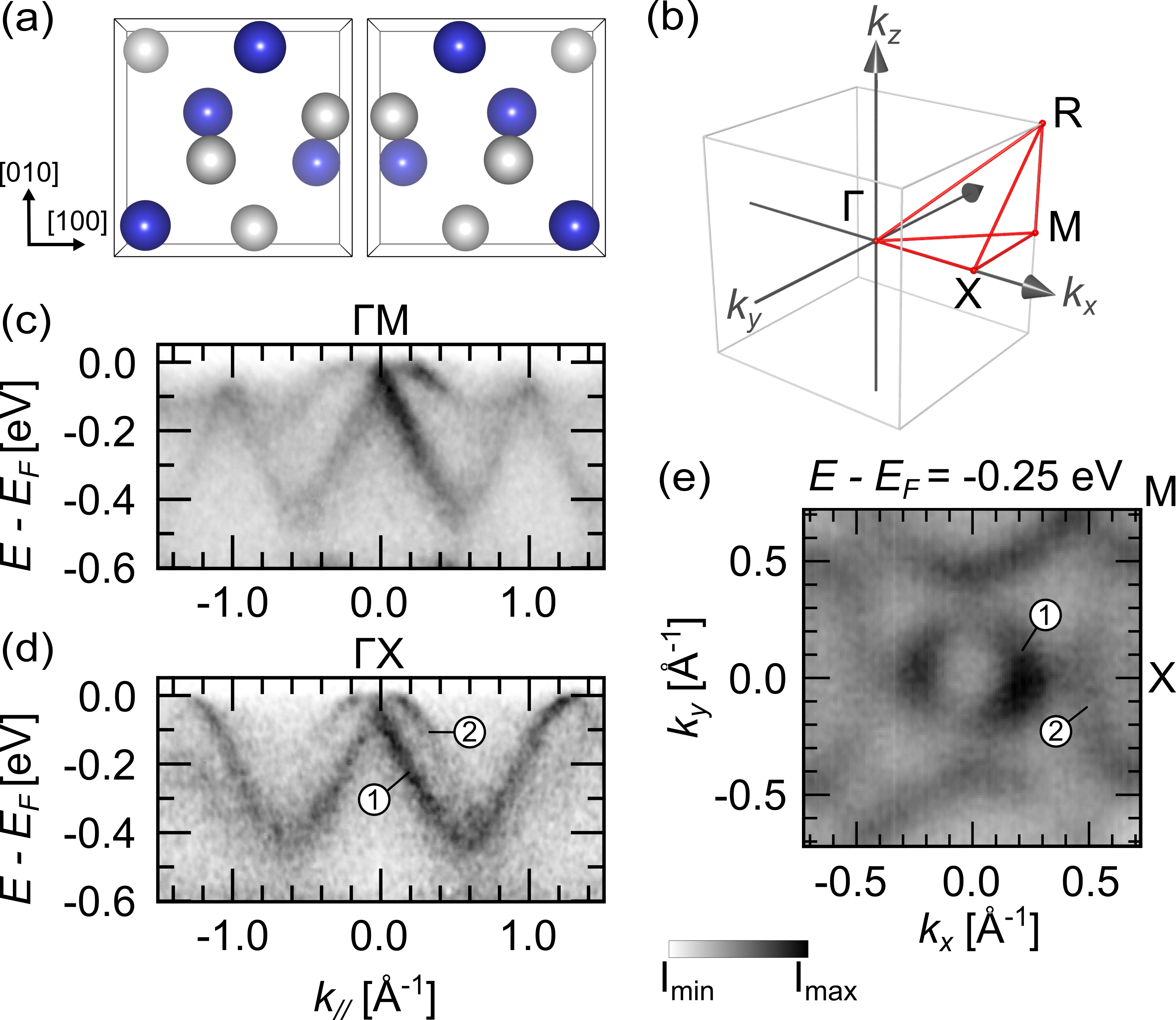}
    \caption{(a) The atomic configuration of enantiomers A and B of CoSi in a unit cell with Co in blue and Si in white. (b) The high-symmetry points in a simple cubic Brillouin zone. ARPES data sets for CoSi(001) along the high-symmetry lines (c) $\Gamma$-$\mathrm{M}$ and (d) $\Gamma$-$\mathrm{X}$, obtained for enantiomer A at a photon energy of $h\nu =$~593~eV. (e) Constant-energy momentum distribution in the $\Gamma$-$\mathrm{X}$-$\mathrm{M}$ plane. Labels (1) and (2) indicate two branches of the multifold band crossing at the $\Gamma$ point \cite{Tang2017, Rao2019}.}
    \label{figure1}
\end{figure}

\begin{figure}[h!]
    \centering
    \includegraphics[width=0.8\linewidth]{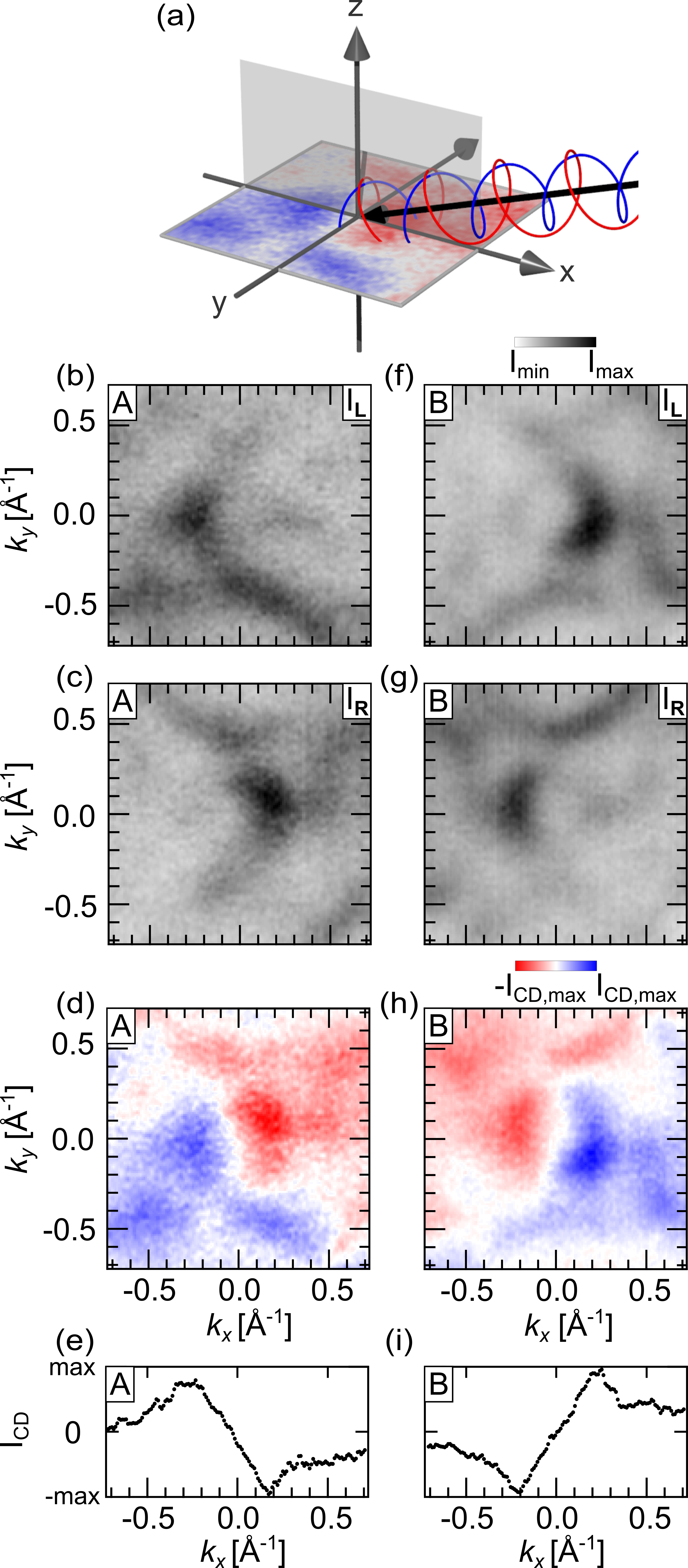}
    \caption{(a) Geometry of the ARPES experiment using incoming left ($I_{L}$) and right ($I_{R}$) circularly polarized light, in the gray $xz$-plane. (b)-(c), (f)-(g) Isoenergy momentum distributions in the $\Gamma$-$\mathrm{X}$-$\mathrm{M}$ plane for enantiomer A (left panels) and enantiomer B (right panels), obtained using $h\nu = 593$~eV and $E-E_{F} = -0.25$~eV. The $\Gamma$-$\mathrm{X}$ direction is aligned along the plane of light incidence. The momentum distributions were taken with (b), (f) left and (c), (g) right circularly polarized light. (d), (h) show the corresponding circular dichroism, $I_{CD} = I_{L} - I_{R}$, for enantiomer A and B, respectively. (e), (i) Circular dichroism $I_{CD}$ in the plane of light incidence ($k_{y} = 0$) extracted from (d) and (h), respectively.}
    \label{figure2}
\end{figure}

\begin{figure*}[ht!]
    \centering
    \includegraphics[width=0.8\linewidth]{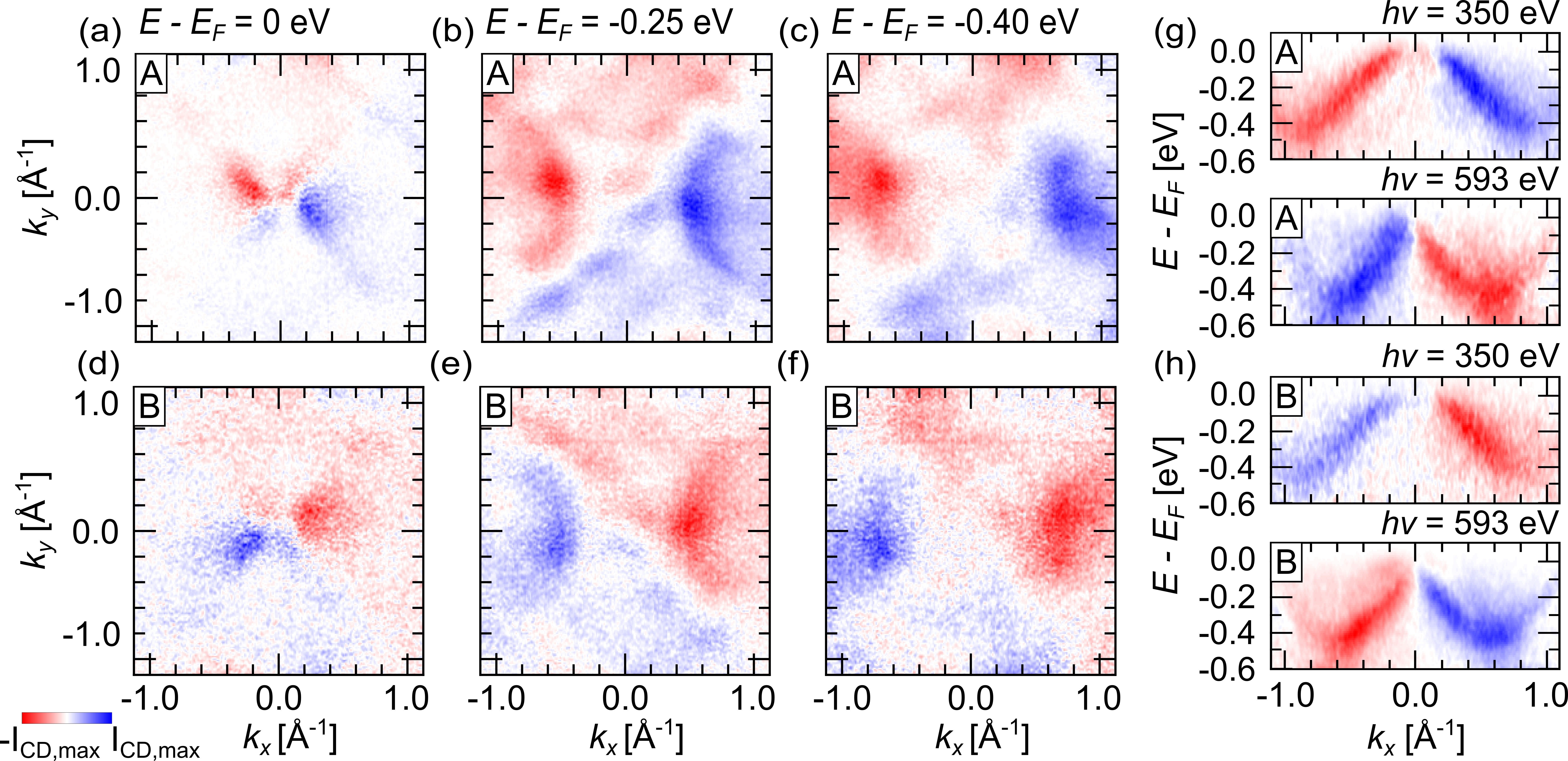}
    \caption{Constant-energy momentum distributions of the circular dichroism in the $\Gamma$-$\mathrm{X}$-$\mathrm{M}$ plane for (a)-(c) enantiomer A and (d)-(f) enantiomer B. The data sets were obtained at a photon energy of $h\nu = 350$~eV and at different binding energies as indicated. (g), (h) Circular dichroism in band-structure maps in the plane of light incidence ($k_{y} = 0$), obtained with $h\nu = 350$~eV and $h\nu = 593$~eV.}
    \label{figure3}
\end{figure*}
CoSi belongs to the B20 crystal family and crystallizes in space group $P2_13$ (No.~198). It is non-symmorphic and chiral in the sense that the space group contains only proper rotations. As a consequence, the crystal exists in two distinct enantiomers, A and B. These are related by spatial inversion, \textit{e.g.} reflection. Figure ~\ref{figure1}(a) illustrates the crystal structure of the CoSi enantiomers A and B, with the corresponding Brillouin zone in Fig.~\ref{figure1}(b). ARPES data sets close to the Fermi level in the $\Gamma$-$\mathrm{X}$-$\mathrm{M}$ high-symmetry plane are shown in Figs.~\ref{figure1}(c)-(e). The data agree with previous DFT calculations \cite{Tang2017, Wilde:22} and ARPES studies of the bulk band structure in CoSi \cite{Rao2019, Takane2019}. In particular, the two lower branches of the topological 3-fold band crossing at the $\Gamma$ point \cite{Tang2017, Rao2019} are resolved and labeled in Figs.~\ref{figure1}(d)-(e).

To study the orbital texture in CoSi, we performed CD-ARPES experiments in the $\Gamma$-$\mathrm{X}$-$\mathrm{M}$ plane of the bulk Brillouin zone. We used photon energies of $h\nu = 350$~eV and $h\nu = 593$~eV as determined from photon energy scans of the $k_z$-dispersion  (see Supplemental Material \cite{suppl}). The circular dichroism (CD) intensity is expressed as $I_{CD} = I_{L} - I_{R}$, where $I_L$ and $I_R$ are the photoemission intensities for left and right circularly polarized photons. Figure~\ref{figure2}(a) illustrates the experimental setup. The photon beam is incident in the $xz$ plane highlighted in gray, at a grazing angle of 20\textdegree. The intensity momentum distributions strongly vary depending on photon polarization and enantiomer type as shown in Figs.~\ref{figure2}(b), (c), (f) and (g). This results in strongly enantiomer-specific CD distributions as shown in Figs.~\ref{figure2}(d) and (h), in qualitative agreement with recent photoemission calculation for chiral crystals \cite{fecher2022}. Interestingly, we observe a strong CD signal in the $xz$ plane of light incidence with a pronounced $\pm k_x$ anti-symmetry, seen in Figs.~\ref{figure2}(e) and (i). Similar characteristics are observed in Figs.~\ref{figure3}(a)-(f), where we show CD momentum distributions for another photon energy and at various binding energies. As we will discuss in more detail below, the dichroic responses in Figs.~\ref{figure2} and~\ref{figure3} reflect the absence of inversion and mirror symmetries in our structurally chiral system. Indeed, the observed momentum distributions are notably different from previous CD-ARPES experiments on non-chiral systems (see, \textit{e.g.}, Refs. ~\cite{Mulazzi:09,Wang:11,Park:2012,bahramy2012,Kim:12, Scholz:13,wiessner2014,Ryu:17,miyamoto2018, Cho:18, cho2021,Fedchenko2019, Unzelmann2021}), where mirror symmetries impose additional constraints on the orbital and CD textures. For example, a crystalline mirror plane aligned with the $xz$ plane of incidence results in vanishing $I_{CD}$ at $k_y = 0$, and $I_{CD}(k_y) = - I_{CD}(-k_y)$ \cite{Park:2012,Kim:12,Scholz2013,Unzelmann2021}. Overall, the enantiomer-specific CD textures in Figs.~\ref{figure2} and~\ref{figure3} indicate chirality-driven orbital textures in CoSi.

As a result of the chiral crystal structure, the Bloch wave functions in CoSi are predicted to carry an OAM texture $\bm{L}(\bm{k})$, with features distinct from non-centrosymmetric but achiral systems. In particular, a monopole-like orbital-momentum locking is expected around the $\Gamma$ and $\mathrm{R}$ points, related to the topological band crossing at these high-symmetry points \cite{Yang2023}. Consistent with this, our CD-ARPES data reveal manifestations of OAM monopoles. To see this, we first focus on the $\Gamma$-$\mathrm{X}$ line, which lies in the plane of light incidence of our experimental setup [Fig.~2(a)]. Along $\Gamma$-$\mathrm{X}$, the crystal symmetries in CoSi impose $\bm{L}(k_x)=L_x(k_x)\hat{e}_x$ with $L_x(-k_x) =-L_x(k_x)$, reflecting the monopole. At the level of the crystal unit cell, the wave functions may be represented as $\ket{\Psi} = \ket{d_{xz}}  + i\gamma(k_x) \ket{d_{xy}}$ with $\gamma(-k_x)=-\gamma(k_x)$ and $L_x \propto \gamma(k_x)$. Here we focus on $d_{xz}$ and $d_{xy}$ orbitals ($l=1$), which appear to be dominant around $\Gamma$ according to calculations \cite{dutta2021}. However, analogous considerations also apply to a combination of $d_{yz}$ and $d_{y^2-z^2}$ ($l=2$). Note that $L_x$ arises from hybridization with an imaginary phase $i\gamma$ between orbitals of even and odd symmetry with respect to the $xz$ plane of light incidence. The electric field vectors of left and right circularly polarized light are approximately $\bm{E}=(0,\pm i\mathcal{E}_y,\mathcal{E}_z)$. Here, for the moment, we neglect the comparably small $\mathcal{E}_x$ component in our grazing-incidence geometry, the role of which will be discussed below. We obtain a relation between OAM and CD in the plane of light incidence via dipole selection rules, assuming an even free-electron final state \cite{moser2017,day2019}. In particular, we find $I_{CD}(k_x)\propto\gamma(k_x)\cdot\Re({T_y^{*} T_z})$, with the transition matrix elements $T_z=\bra{\bm{k}}\mathcal{E}_z p_z\ket{d_{xz}}$ and $T_y=\bra{\bm{k}}\mathcal{E}_y p_y\ket{d_{xy}}$. The results of this simplified model agree with previous results on the relation of CD and OAM \cite{Park2012a,Schueler2020}.

The sizable CD signals in the plane of light incidence demonstrate the formation of OAM $L_x$ parallel to the momentum $k_x$. Furthermore, the measured reversal $I_{CD}(-k_x)=-I_{CD}(k_x)$ reflects the monopole-like reversal of OAM $L_x$ across the $\Gamma$ point, since both originate from the sign change in the complex phase $i\gamma$ between different $d$ orbital components. Going from enantiomer A to enantiomer B, the OAM monopole at $\Gamma$ is predicted to transform into an anti-monopole \cite{Yang2023}. Within our simplified consideration, this corresponds to a sign change $\gamma_B(k_x)=-\gamma_A(k_x)$. Indeed, our experimental data show $I_{CD}^{A}(k_x)=-I_{CD}^{B}(k_x)$ in the plane of light incidence [see, \textit{e.g.}, Figs.~\ref{figure2}(e) and (i)]. This demonstrates an OAM reversal between the two enantiomers. 

Figures \ref{figure3}(g) and \ref{figure3}(h) show CD-ARPES data sets along $\Gamma$-$\mathrm{X}$ for the two photon energies $h\nu = 350$~eV and $h\nu = 593$~eV. The CD sign changes in the plane of light incidence, \textit{i.e.}, $I_{CD}(k_x) = -I_{CD}(-k_x)$ within enantiomers, and $I_{CD}^{A}(k_x) = -I_{CD}^{B}(k_x)$ between enantiomers, are persistent over the full band width. Interestingly, for a given enantiomer, the CD signal changes sign between the two photon energies. Within the model discussed above, this sign change can be attributed to the term $\Re({T_y^{*} T_z})$, since the matrix elements $T_y$ and $T_z$ depend on the final state and therefore vary with photon energy. Similar effects have been observed for dichroic and spin-resolved intensities in two-dimensional systems with spin- and orbital-polarized states \cite{Scholz:13,Bentmann:17,Bentmann:21}. 

We now consider the symmetry properties of the CD momentum distributions in Figs.~\ref{figure2} and \ref{figure3} in more detail. The experimental data show that (i) $I_{CD}(k_x,k_y) = -I_{CD}(-k_x,-k_y)$. This relation can be understood based on the $C_{2z}$ crystal symmetry of CoSi, which relates $(k_x,k_y)$ and $(-k_x,-k_y)$. Furthermore, assuming fully grazing incidence, $C_{2z}$ swaps the light polarization state from left to right circular, \textit{i.e.}, $(0,+ i\mathcal{E}_y,\mathcal{E}_z)$ to $(0,-i\mathcal{E}_y,\mathcal{E}_z)$, so that $I_{R}(k_x,k_y)= I_{L}(-k_x,-k_y)$. Comparing the data sets for different enantiomers in Figs.~\ref{figure2} and \ref{figure3}, we observe (ii) $I_{CD}^{A}(k_x,k_y)= I_{CD}^{B}(-k_x,k_y)$ and (iii) $I_{CD}^{A}(k_x,k_y) = -I_{CD}^{B}(k_x,-k_y)$. These relations are connected to the mirror operation $\mathcal{M}_x$, which transforms the crystal structure of the two enantiomers into each other. The light polarization $(0,\pm i\mathcal{E}_y,\mathcal{E}_z)$ is invariant under $\mathcal{M}_x$ such that $I_{R,L}^{A}(k_x,k_y)= I_{R,L}^{B}(-k_x,k_y)$. Relation (iii) can be understood as a direct consequence of the relations (i) and (ii).

The finite $\mathcal{E}_x$ component of the light in our experimental geometry could be expected to induce deviations from the above relations (i) and (ii). The role of $\mathcal{E}_x$ may be grasped as an additional contribution to the CD textures that is even under the $C_{2z}$ operation and arises from a light field component $(\mathcal{E}_x,\pm i\mathcal{E}_y,0)$. However, such deviations are marginal in our data sets at both photon energies. Due to the grazing light incidence, this contribution is attenuated for geometrical reasons. Moreover, in the plane of light incidence $(\mathcal{E}_x,\pm i\mathcal{E}_y,0)$ is expected to couple to the $L_z$ OAM component, which vanishes in the $\Gamma$-$\mathrm{X}$-$\mathrm{M}$ plane \cite{Yang2023}. In this regard, photon energy-dependent CD-ARPES measurements in a geometry closer to normal incidence could be an interesting approach to address the $L_z$ component across the OAM monopole.   

\begin{figure}[t!]
    \centering
    \includegraphics[width=0.8\linewidth]{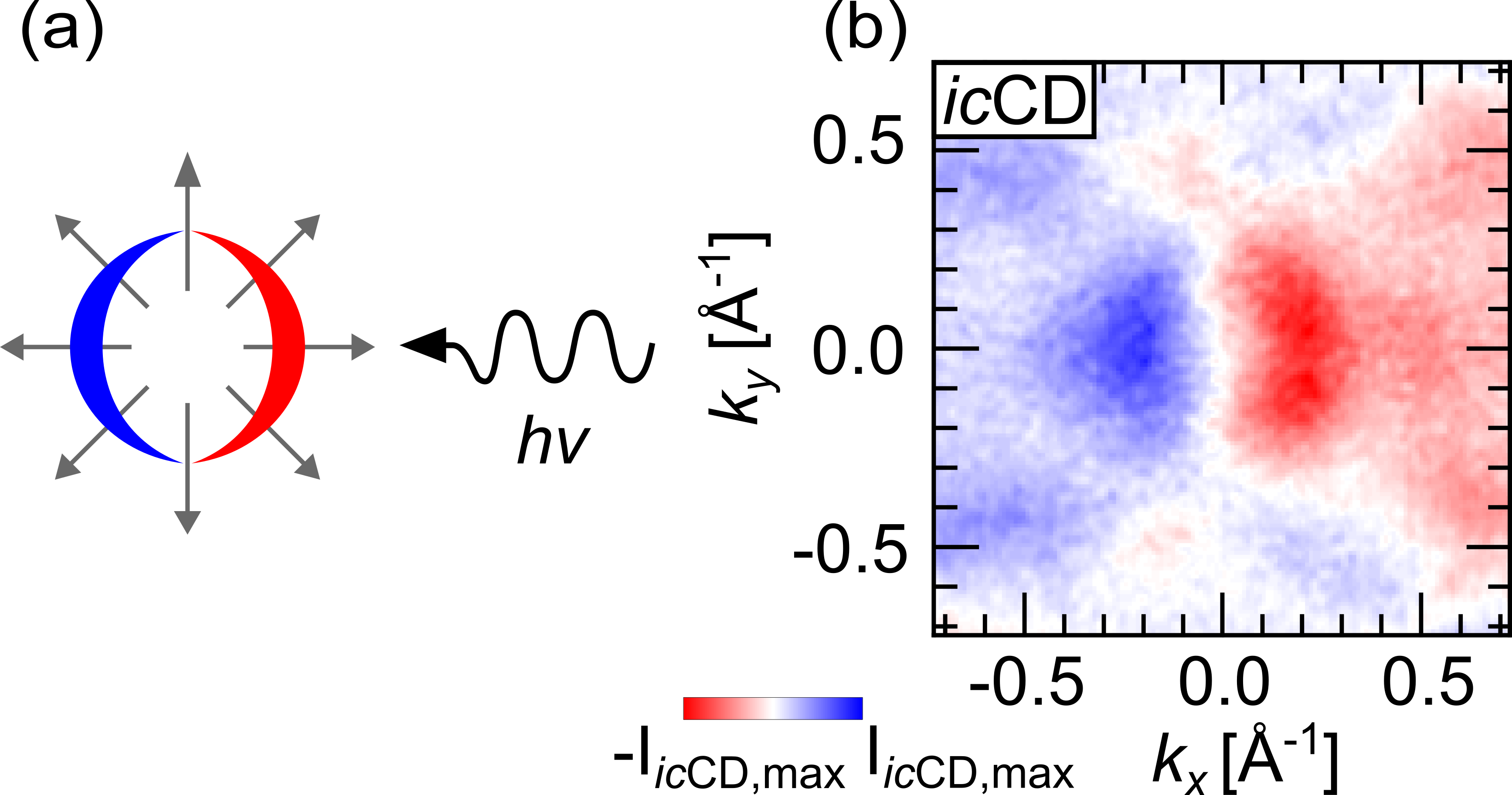}
    \caption{(a) Schematic of the OAM monopole (gray arrows) in the $\Gamma$-$\mathrm{X}$-$\mathrm{M}$ plane according to Ref.~\cite{Yang2023}, and the corresponding dichroic signal. (b) Momentum distribution of the intrinsic chiral circular dichroism, $ic\mathrm{CD}=\frac{1}{2}(I_{CD}^{A}-I_{CD}^{B})$, at $h\nu = 593$~eV and $E-E_{F} = -0.25$~eV.}
    \label{figure4}
\end{figure}

Let us consider manifestations of the predicted OAM monopole \cite{Yang2023} in the measured CD texture beyond the plane of light incidence discussed above. While the CD signal generally reflects the OAM $L_x$ texture \cite{Park2012a}, it may also contain extrinsic contributions that arise from the experimental geometry-induced symmetry breaking \cite{schonhense1990} (see Supplemental Material \cite{suppl}). Considering the data sets $I_{CD}^{A}(k_x,k_y)$ and $I_{CD}^{B}(k_x,k_y)$ in Figs.~\ref{figure2} and \ref{figure3}, we notice that the main features change sign between enantiomers while no or only weak changes are observed in other momentum regions. The latter may be understood by the fact that contributions to the CD signal originating from experimental geometry remain unchanged upon the exchange of enantiomer. We introduce the \textit{intrinsic chiral} circular dichroism which we define as $ic\mathrm{CD}=\frac{1}{2}(I_{CD}^{A}-I_{CD}^{B})$. The approach is inspired by Refs.~\cite{beaulieu:2021,Beaulieu:21}, which considered a differential dichroic signal resulting from two different orientations of one achiral crystal. By contrast, we compare data sets for two different crystals in the same orientation, the two enantiomers A and B. The $ic\mathrm{CD}$ allows us to extract contributions to $I_{CD}$ arising from structural chirality, which imposes equivalent band structures but inequivalent OAM textures for the two enantiomers. In Fig.~4 (and in the Supplemental Material \cite{suppl}), we consider the $ic\mathrm{CD}$ signal, which is in good agreement with a monopole-like $L_x$ texture across the full momentum distribution. Moreover, we find that, to good approximation, $ic\mathrm{CD}(k_x,k_y)$ shows the expected symmetry properties of the $L_x$ component, namely $L_x(k_x,k_y)=-L_x(-k_x,k_y)$. The $ic\mathrm{CD}$ thus confirms key characteristics of the chirality-driven OAM texture.   

Our experiments exploit the increased probing depth in the soft X-ray photon energy regime to probe the bulk electronic chirality and orbital texture. We anticipate that more surface-sensitive CD-ARPES in the vacuum ultraviolet regime could provide an interesting approach to exploring the interplay of surface and bulk states in chiral topological semimetals \cite{Min:19}. The pronounced handedness-dependence of the photoemission response we observe in CoSi also motivates spin-resolved measurements, particularly in view of possible CISS effects in photoemission \cite{gohler2011}. 
In contrast to the photoelectron CD observed in ensembles of randomly-oriented chiral molecules in the gas phase, typically on the order of a few percent \cite{Heinzmann:01, Beaulieu2016}, we observe here a markedly higher chiral CD response on the order of the absolute intensity. In view of uncoventional excited-state dynamics observed in chiral molecules upon exciation with circularly polarized light pulses \cite{Beaulieu2018}, this substantial enhancement in CD underscores the potential of chiral crystals in the generation and manipulation of orbital photocurrents \cite{Adamantopoulos2024}.

In conclusion, we studied the orbital texture in the bulk electronic structure of CoSi using circular dichroism in bulk-sensitive soft X-ray ARPES. Our measurements unveil a strong and highly enantiomer-specific dichroic response, demonstrating how electronic chirality arises from the handedness of the chiral crystal structure. Our experimental findings confirm recent theoretical predictions of monopole-like OAM textures around high-symmetry points in chiral topological semimetals \cite{Yang2023}. This exotic OAM texture is expected to result in transport phenomena, such as orbital current generation \cite{Yang2023}, and highlights the potential for applications of chiral topological semimetals in spin-orbitronics. Furthermore, our observation of chirality-driven OAM textures support recent theory predicting an OAM-assisted origin of the CISS effect in chiral structures \cite{liu:2021}.

We have introduced the intrinsic chiral circular dichroism, \textit{ic}CD, as a differential photoemission observable sensitive to electronic chirality. We envision that the \textit{ic}CD may enable new types of experiments for a variety of systems with chiral electron states. Besides crystalline and molecular materials with structural chirality, these may include chiral magnets \cite{wilde2021}, and electronic-instability-driven states with chiral charge order, \textit{e.g.}, in Kagome metals \cite{jiang2021}, or chiral charge-density waves \cite{Yang22}.  

Note: After completion of this work, a preprint on CD-ARPES experiments on a different chiral topological semimetal became available \cite{yen2023controllable}.
\\
\\
\begin{acknowledgments}
HB thanks Maximilian \"Unzelmann for insightful discussions. This work was supported by the Research Council of Norway through Grant No. 323766 and its Centres of Excellence funding scheme Grant No. 262633 “QuSpin.” We acknowledge DESY (Hamburg, Germany), a member of the Helmholtz Association HGF, for the provision of experimental facilities. Parts of this research were carried out at PETRA III and we would like to thank Jens Viefhaus, Frank Scholz, and J\"orn Seltmann and for assistance in using beamline P04. Funding for the photoemission spectroscopy instrument at beamline P04 by the Federal Ministry of Education and Research (BMBF) is gratefully acknowledged (Contracts 05KS7FK2, 05K10FK1, 05K12FK1, 05K13FK1, 05K19FK4 with Kiel University; 05KS7WW1, 05K10WW2 and 05K19WW2 with the University of W\"{u}rzburg). Beamtime was allocated for proposal II-20230019.
\end{acknowledgments}

\bibliography{refs.bib}

 

\widetext
\clearpage
\begin{center}
\textbf{\large Supplemental material: Chirality-Driven Orbital Angular Momentum and Circular Dichroism in CoSi}
\end{center}
\setcounter{equation}{0}
\setcounter{figure}{0}
\setcounter{table}{0}
\setcounter{page}{1}

\onecolumngrid

\section*{Sample characterizations}
\subsection*{Sample growth}

\begin{figure}[h]
    \centering
    \includegraphics[width=0.8\linewidth]{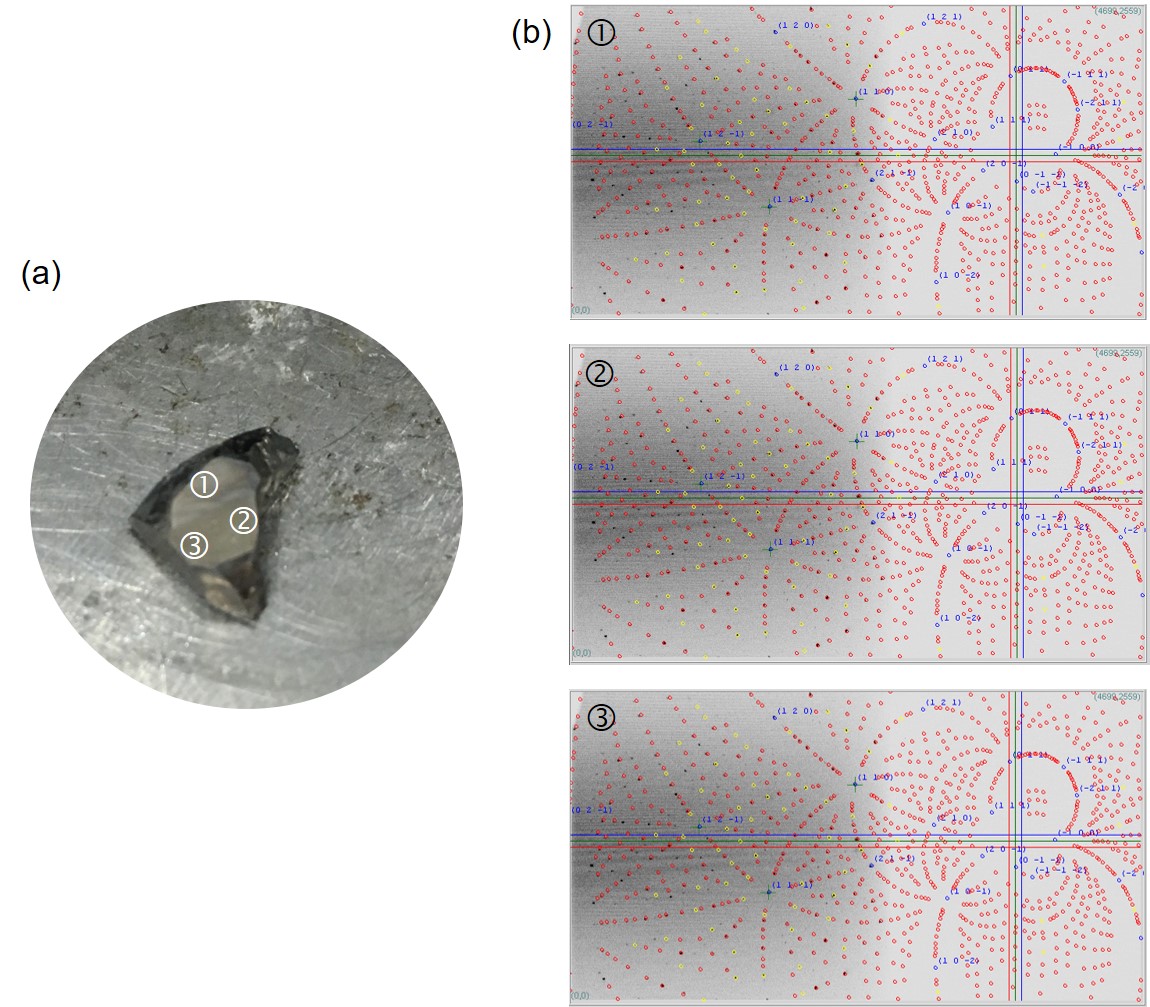}
    \caption{(a) The morphology of CoSi single crystal and (b) Laue-XRD diffraction patterns at different positions, marked in (a), on the surface.}
    \label{Laue}
\end{figure}

To grow CoSi single crystals, we first prepare the polycrystalline CoSi powders by mixing Co elemental powder (99.995\%, ACROS) with Si elemental powder (99.995\%, ACROS) in a molar ratio of 1:1. Then, the mixture was heated at 1200°C withholding this temperature for 48 hours for high-temperature solid-state synthesis. During this step, hydrogen gas (H$_2$) was injected to prevent surface oxidation of Si which could inhibit the progress of the synthesis reaction. In this study, we employed the Czochralski technique to grow high-quality CoSi single crystals \cite{Marshall1968}. This method allows to select a specific crystallographic plane of a CoSi single crystal as a seed crystal and further growth of the desired crystal faces. To prevent the formation of oxides (\textit{e.g.}, SiO$_2$) during the single crystal growth process, the atmosphere was maintained at a vacuum level of 10$^{-5}$ Torr. The prepared CoSi single crystal (as shown in Fig. \ref{Laue}(a)) was characterized using Laue-XRD with random sampling points to confirm that the entire measured surface belonged to the (100) crystal plane after polishing as a mirror condition. Finally, electropolishing was employed to remove residual stress layers on the surface, reducing errors introduced by stress fields during measurements.

\subsection*{Surface preparation}
We prepared clean (001) surfaces of the two CoSi enantiomers using cycles of sputtering with $\mathrm{Ar}^+$ at an energy of 2 keV, followed by annealing at $\sim$950~K to restore single crystallinity. To evaluate the preparation method, we regularly performed low energy electron diffraction (LEED) experiments. Sharp dots indicate a high surface quality and confirm the crystal orientation. 
A two-fold $C_{2z}$ symmetry can be observed in the LEED patterns. The inversion of the LEED patterns between enantiomers A and B shows the opposing chirality of the crystals. 

\begin{figure}[h!]
    \centering
    \includegraphics[width=0.65\linewidth]{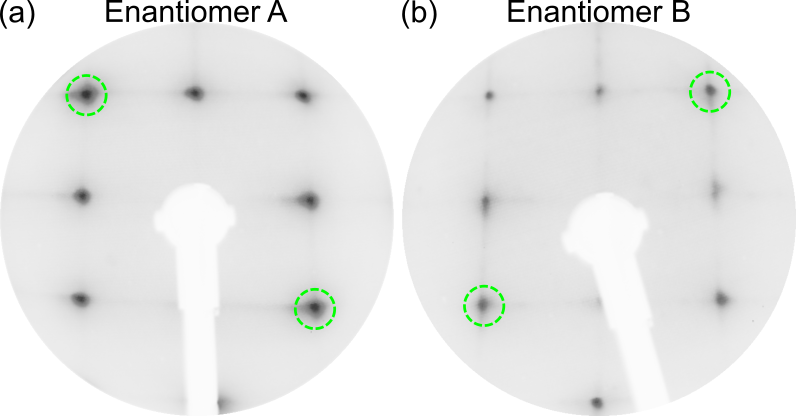}
    \caption{Low energy electron diffraction (LEED) of (a) enantiomer A and (b) enantiomer B obtained at 60~eV electron energy.}
    \label{LEED}
\end{figure}

\section*{Out-of-plane momentum dependence}
\begin{figure}[h]
    \centering
    \includegraphics[width=0.6\linewidth]{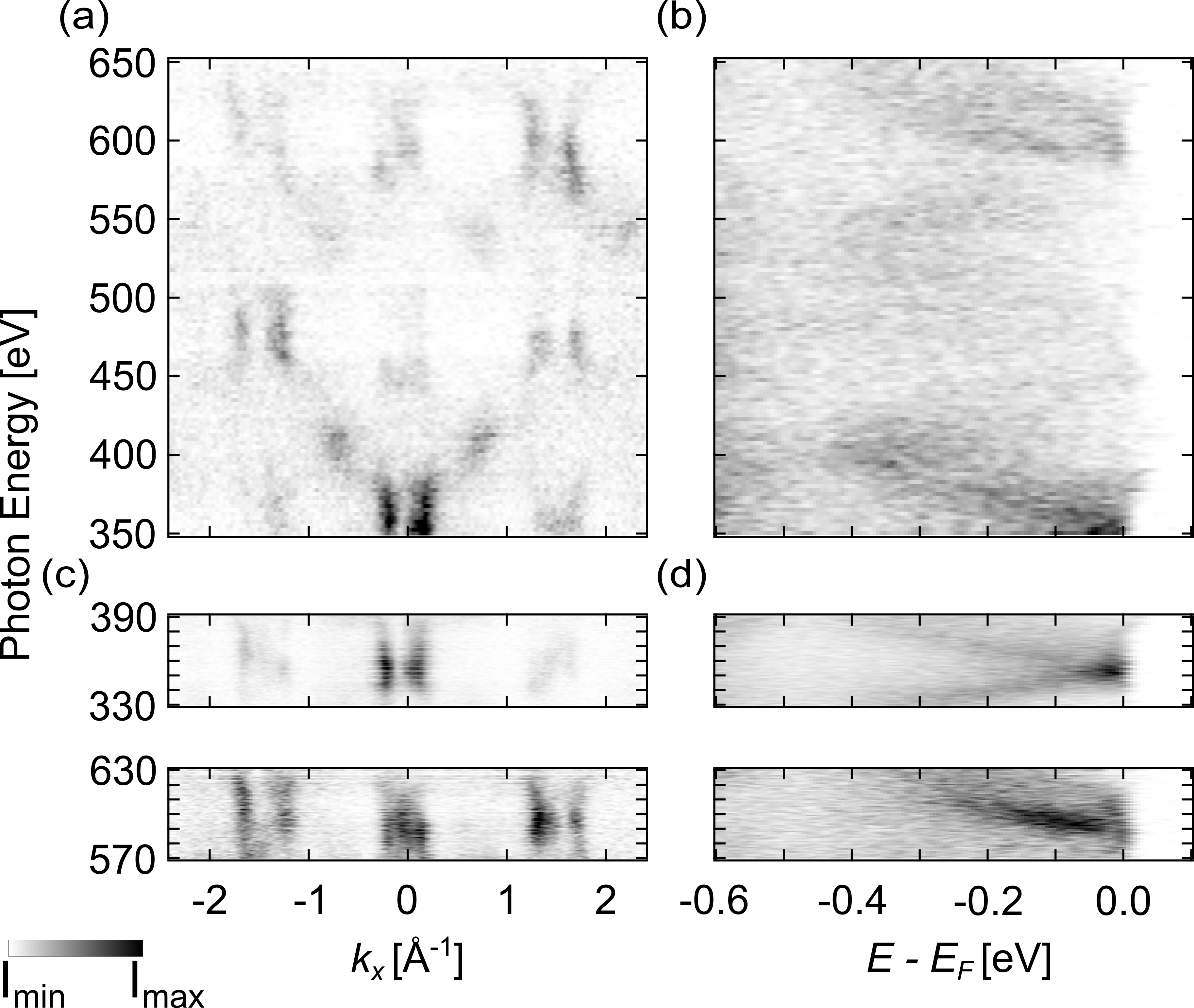}
    \caption{Photon energy dependent ARPES of CoSi. The $\Gamma$-$\mathrm{X}$ direction is aligned in the plane of light incidence. (a) and (c) show the photon energy dependence on $k_x$ at $E - E_{F} = 0$ eV. (a) contains an overview of an extended range of soft X-ray photon energies, whereas (c) shows photon energies focused around the two $\Gamma$ points. (b) and (d) depict the corresponding band dispersions as a function of photon energy.}
    \label{hvscan}
\end{figure}

The main text considers the photoemission response of CoSi using two distinct photon energies $h\nu = 350$ eV and $h\nu = 593$ eV. Here, we demonstrate that these two photon energies probe the $\Gamma$ point, by mapping out the ARPES photoelectron intensity along the $\Gamma$-$\mathrm{X}$ line as a function of photon energies $h\nu$. Figure \ref{hvscan} shows an overview (a), (b) and a more detailed view (c), (d) of the obtained photon energy scans in the soft X-ray range. Only a single light polarization was used. In Fig. \ref{hvscan}(a), one can see the periodic recurrence of the $\Gamma$ point as a function of photon energy, which is related to the out-of-plane momentum $k_\perp$. In particular, from Fig. \ref{hvscan}(c) it can be seen that with our choice of $h\nu = 350$ eV, we probe a $\Gamma$ point in CoSi, in line with Ref. \cite{Rao2019}. Using $h\nu = 593$ eV, the $\Gamma$ point lying in a different Brillouin zone is measured.

\section*{Intrinsic chiral circular dichroism}
\begin{figure}[h]
    \centering
    \includegraphics[width=0.3\linewidth]{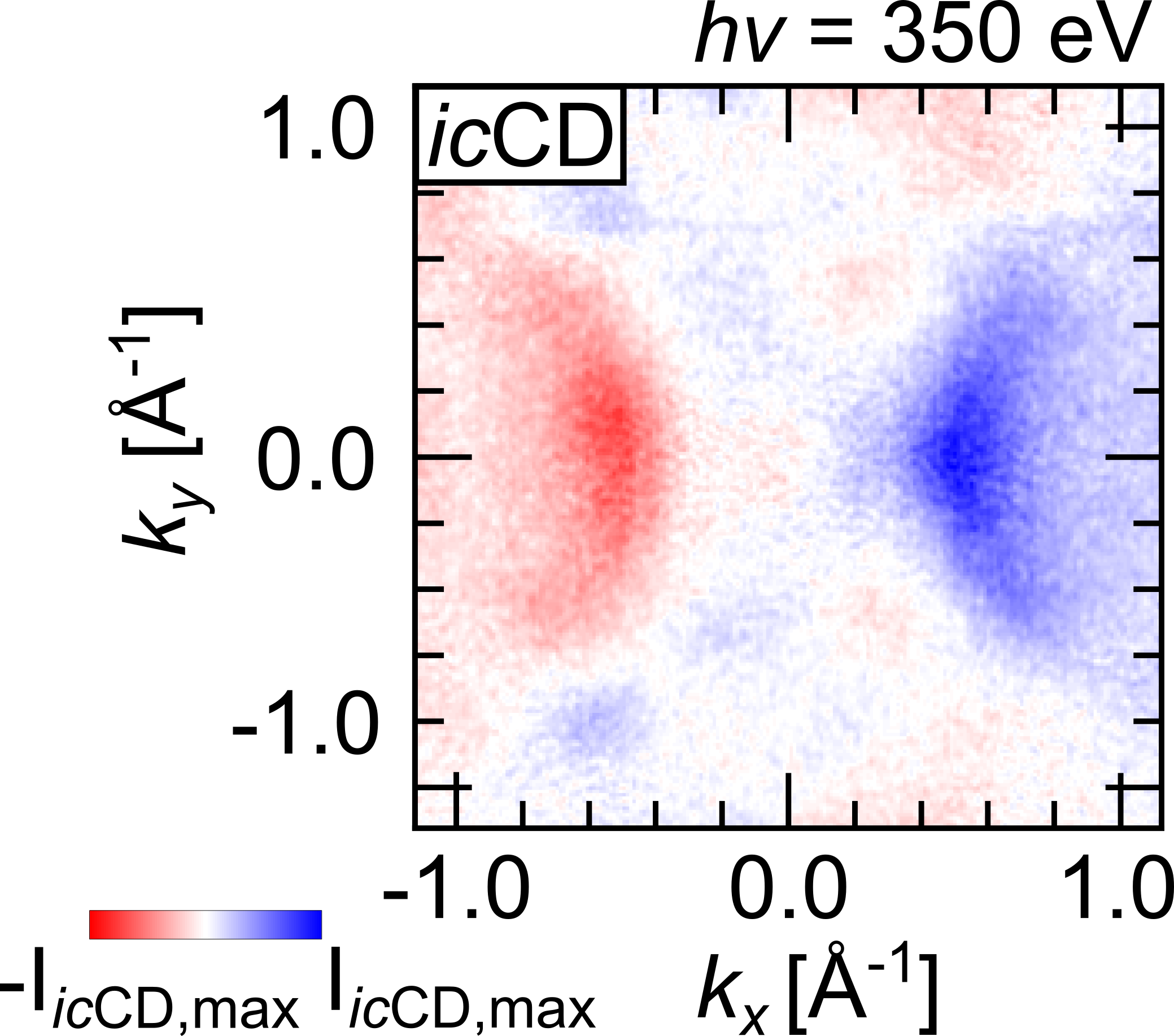}
    \caption{Momentum distribution of the intrinsic chiral circular dichroism, $ic\mathrm{CD}=\frac{1}{2}(I_{CD}^{A}-I_{CD}^{B})$, at $h\nu = 350$~eV and $E-E_{F} = -0.25$~eV.}
    \label{iccd_350}
\end{figure}

In the main text, we introduced the \textit{intrinsic chiral} circular dichroism, $ic\mathrm{CD}=\frac{1}{2}(I_{CD}^{A}-I_{CD}^{B})$, as a photoemission observable. In the $ic\mathrm{CD}$, extrinsic contributions induced by the experimental geometry should be diminished compared to the CD texture. Instead, we emphasize contributions originating from structural chirality. Figure \ref{iccd_350} shows the $ic\mathrm{CD}$ signal for a different photon energy than shown in the main text. We observe similar features as seen previously (\textit{cf.} Fig. 4 in the main text): again, the $ic\mathrm{CD}$ signal is in line with a monopole-like $L_x$ texture of the OAM.

\section*{Non-chiral contributions to the circular dichroism}
\begin{figure}[h]
    \centering
    \includegraphics[width=0.3\linewidth]{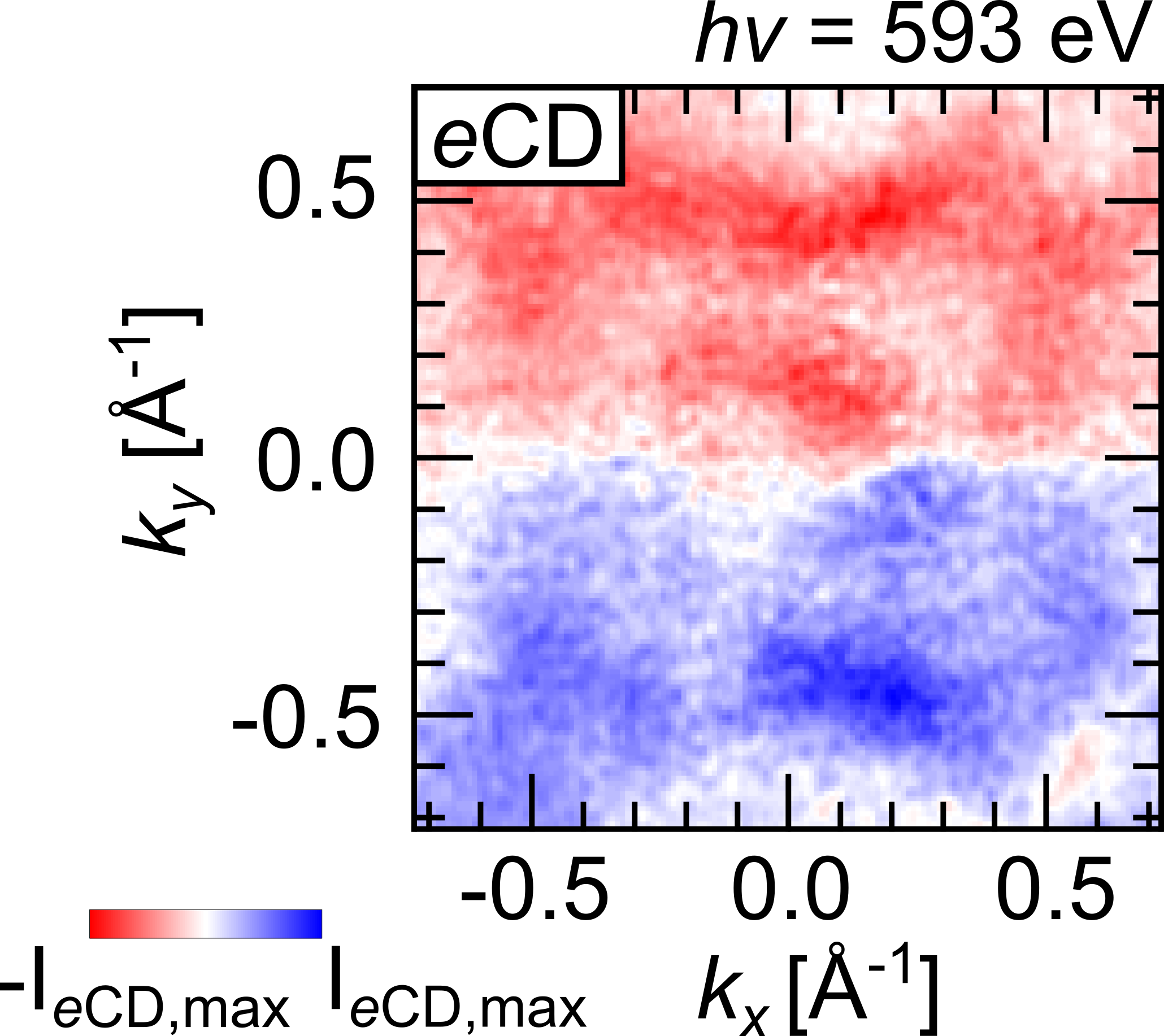}
    \caption{Momentum distribution of the extrinsic contributions to the circular dichroism, $e\mathrm{CD}=\frac{1}{2}(I_{CD}^{A} + I_{CD}^{B})$, at $h\nu = 593$~eV and $E-E_{F} = -0.25$~eV.}
    \label{ecd_593}
\end{figure}

In the $ic\mathrm{CD}$, non-chiral or ``extrinsic'' contributions to the dichroic response induced by the experimental geometry are diminished. To gain more insight into these extrinsic contributions, we consider the sum of the dichroic response of enantiomers A and B, $e\mathrm{CD}=\frac{1}{2}(I_{CD}^{A} + I_{CD}^{B})$, shown in Fig. \ref{ecd_593}. For the $e\mathrm{CD}$ signal, we expect the influence of structural chirality on the CD signal to be suppressed. Indeed, it can be readily seen that the expected response for a mirror-symmetric systems is regained, namely a vanishing CD in the plane of light incidence $I_{CD}(k_y = 0) = 0$, and an anti-symmetry with respect to the $k_y = 0$ axis: $I_{CD}(k_y) = - I_{CD}(-k_y)$.
 
To gain a simple picture of the $e\mathrm{CD}$, we may consider a purely $s$-like band that does not ``feel'' the chiral crystal potential. In this case, we expect the $ic\mathrm{CD}$ to be largely suppressed. However, a finite $e\mathrm{CD}$ signal is still expected based on the mirror-symmetry breaking of the circular light polarization and the superposition of different final-state partial waves \cite{Fedchenko2019}. For a state influenced by the structural chirality, the total CD-ARPES response is composed by the sum of $ic\mathrm{CD}$ and $e\mathrm{CD}$. 

\section*{Photon energy dependence of the circular dichroism}
In Figure 3(g)-(h) of the main text, we report a sign reversal of the circular dichroism in the plane of light incidence as a function of photon energy. 
This observation can be attributed to different final-state wave functions for the two photon energies. In the main text, we discuss a simple model that allows us to qualitatively capture the influences of the initial-state OAM and the final state on the CD signal. In the plane of light incidence, we have $I_{CD}(k_x)\propto\gamma(k_x)\cdot\Re({T_y^{*} T_z})=\gamma(k_x)\Re{(|T_y^{*}|\cdot| T_z|\exp(i(\phi_z-\phi_y))}$. The handedness- and momentum-dependence are captured by the phase $\gamma(k_x)$ which is related to the initial-state OAM. The phases $\phi_z$ and $\phi_y$ of the photoemission matrix elements depend on the final-state wave function, which can result in a photon energy dependence of the CD signal, as observed experimentally. We note here that similar photon energy dependencies of the CD signal and the spin-resolved photoemission signal for spin- and orbital-polarized states have been observed previously and were shown to arise from variations of the final-state wave function with energy \cite{Scholz:13,Bentmann:17,Bentmann:21}. A more detailed understanding of the absolute sign of the CD signal in CoSi will require advanced photoemission calculations \cite{Scholz:13,Bentmann:21}.

\end{document}